# Using Thermodynamics and Microstructure to Mitigate Overfitting in Pellet Reduction Models

*Ömer K. Büyükuslu*, Fabrice Yang, Dierk Raabe, Moritz to Baben, and Anna L. Ravensburg*

Ö. K. Büyükuslu, F. Yang, M. to Baben, A. L. Ravensburg
GTT-Technologies
Kaiserstraße 103, 52134 Herzogenrath, Germany
E-mail: ob@gtt-technologies.de

Ö. K. Büyükuslu, D. Raabe
Max Planck Institute for Sustainable Materials
Max-Planck-Straße 1, 40237 Düsseldorf, Germany

## Abstract

Direct reduction of iron using hydrogen-rich gas is rapidly emerging as a key strategy for green steel production. This process involves complex, multiscale phenomena, encompassing solid-state phase transformations and gas transport through pores, that must be accurately represented for predictive industrial implementation. Here, we present a thermodynamically sound pellet-scale model that describes these mechanisms and can serve as a foundation for improving the understanding of pellet reduction kinetics in $H_2/CO$-containing atmospheres. The model assumes that the gas phase remains in thermodynamic equilibrium, meaning that the composition of the gas instantaneously adjusts to any changes in the system. This reduces the number of fitting parameters drastically compared to other existing models, while maintaining a strict thermodynamic upper bound estimate. A driving force term is included in the reaction rate equation based on the partial pressure of $O_2$ in the equilibrated gas phase. This constrained equilibrium-based approach ensures that the three iron oxide reduction steps ($Fe_2O_3 \leftrightarrow Fe_3O_4$, $Fe_3O_4 \leftrightarrow FeO$, $FeO \leftrightarrow Fe$) and the formation of graphite and cementite in carbon-containing gases occur only if they are thermodynamically possible. It is demonstrated that fitting kinetic parameters based on conversion degree data alone leads to overfitting. This is true both for existing models and the model introduced here, despite the fact that the latter contains fewer parameters. To overcome this overfitting problem, spatially resolved microstructural data at key reduction stages can be considered, as shown here for recently reported data for a pellet reduced in $H_2$ atmosphere.

## 1. Introduction

The iron and steel industry accounts for 7–9% of anthropogenic global greenhouse gas emissions, making it a major contributor to climate change[1–4]. To address this challenge, the industry is shifting toward sustainable alternatives to reduce greenhouse gas emissions from the initial step of steel production, i.e., producing iron from its ores. Here the Direct Reduction of Iron (*DRI*)-Electric Arc Furnace route stands out as a particularly promising option, as indicated by techno-economic assessments[5–7]. *DRI* is a solid-state reduction process mainly conducted in vertical shaft furnaces, where hematite ($Fe_2O_3$) pellets are charged from the top and reducing gas mixtures are introduced from the bottom, creating a counter-current flow under controlled conditions[8]. When hydrogen is utilized as a reducing agent in this process, it generates water vapor ($H_2O$) as a byproduct, which can then be recycled through the natural water cycle. This approach has the potential to reduce greenhouse gas emissions from approximately 1.8 to 0.2–0.5 t of $CO_2$/t of steel, assuming the adoption of green electrolysis plants to supply hydrogen for future *DRI* facilities[9]. On the other hand, the conventional Blast Furnace-Basic Oxygen Furnace method remains the most $CO_2$-intensive and energy-demanding process in iron and steelmaking[4,10]. Using $H_2$ rather than $CO$ as reductant leads to fundamental changes, e.g., making the process endothermal rather than exothermal[11], changes in the position of equilibria[9] and increasing reduction kinetics. Therefore, it is evident that physically sound process models of *DRI* technology are needed to deepen the understanding of process dynamics, quantify key variable interdependencies, and predict outcomes under variable conditions, such as changing $H_2/CO$ ratios in the reducing gas. Given the economic and logistical challenges associated with hydrogen supply, optimizing the *DRI* process with regards to structural parameters of the pellet (i.e., composition, size, porosity, microstructure) and regarding the process parameters (i.e., temperature, pressure, pellet and gas flow rate, gas composition) is crucial to guarantee a viable cost and energy efficiency.

Modeling *DRI* is challenging since the reduction process in shaft furnaces involves physical and chemical mechanisms on diverse length scales, including phase transformations, nucleation and growth, reaction kinetics, heat and mass transport, and volume changes within the pellets[12,13]. For capturing these multi-scale phenomena, a hierarchical modeling approach is suitable[14], i.e., having distinct pellet-scale and furnace-scale models. In this approach, it is essential to have a detailed, physically sound pellet-scale model which reliably predicts the microstructural evolution in the pellets and enables a robust furnace-scale model through extrapolation. Recently, Salucci et al. critically reviewed modeling approaches on the pellet scale and concluded that there are extensive gaps in the fundamental mechanisms leading to questionable model parametrization[9]. The most widely applied structural pellet models in the literature are the Shrinking Core Model (*SCM*) and the Grain Model (*GM*)[15–18].

The *SCM* is based on the assumption that the reduction of iron oxide pellets can be described by a topochemical structure, i.e., as a shrinking unreacted core with three inward-moving boundaries representing phase transformations from hematite to magnetite ($Fe_3O_4$), wüstite ($FeO$), and finally to iron ($Fe$)[15,17]. The *GM* adds to this and assumes the pellet to be composed of dense, uniformly sized small grains, each of which follow the shrinking core behavior[16,18,19]. Reaction kinetics in these models are defined by the shrinking rate of reaction zones over time, using a rate equation based on the Langmuir-Hinshelwood kinetic expression[20]. This equation incorporates an equilibrium constant $K_{eq}$ that varies with temperature, reflecting thermodynamic stability by assuming local equilibria[17,20–22]. As $K_{eq}$ is independently specified for each reaction equation, these models require defining up to ten parameters in systems utilizing $CO$ and $H_2$ as reducing agents[13,23]. This reliance on that many parameters is a drawback, as each requires extensive experimental validation, especially in systems with more gas species and occurring partitioning reactions. Furthermore, although the *SCM* is a practical pellet-scale model, the assumption of sharp reaction interfaces oversimplifies microstructural evolution, disregarding the influence of structural parameters such as grain size, porosity, and pore size[19]. The *GM* addresses this limitation by incorporating structural parameters as chemical resistances in the reaction equations but still assumes sharp reaction interfaces within each grain and a uniform grain size distribution[11,19]. *SCM* and *GM*, relying on moving reaction fronts, structurally conflict with recent microstructural studies showing significant local variations in grain size, porosity, and phase transformations during reduction[24–29]. These limitations highlight the need for a model resolving time dependent changes across different locations within the pellet, allowing for a more physically sound simulation of the reduction process. Fradet et al. proposed a one-dimensional (*1D*) porous solid model (*PSM*) that solves the governing differential equations originating from mass conservation across the radius of a pellet, assuming spherical symmetry[30]. Unlike the models based on moving reaction fronts, this approach uses spatial discretization employing the Finite Difference Method (*FDM*) to approximate the microstructure at discrete radial grid points[31]. The applied reduction kinetics for solid-gas reactions are described by an irreversible first-order kinetic expression, which, however, does not account for equilibrium dynamics. This poses a limitation, resulting in thermodynamic inconsistencies under conditions where phase formation is thermodynamically limited, as shown in Figure 1. It can be seen in the phase diagram displayed in Figure 1a that at 850°C the formation of metallic iron is thermodynamically possible only for $H_2/(H_2+H_2O)$ ratios larger than 0.62. Below that, *FeO* is more stable, and oxygen cannot be transferred to the gas phase. In Figure 1b the radially resolved phase formation at 50 % global conversion is displayed predicted in the *PSM* by Fradet et al., where metallic Fe is observed at radii larger than 3 mm, while the $H_2/(H_2+H_2O)$ ratio becomes larger than 0.62 only above radii larger than 4 mm. While kinetic limitations can lead to slow decomposition rates, kinetic limitations cannot lead to phases forming from more stable reactants. Therefore, a modification of the *PSM* model is needed to include thermodynamic consistency while maintaining the possibility to model the radially resolved pellet microstructure.

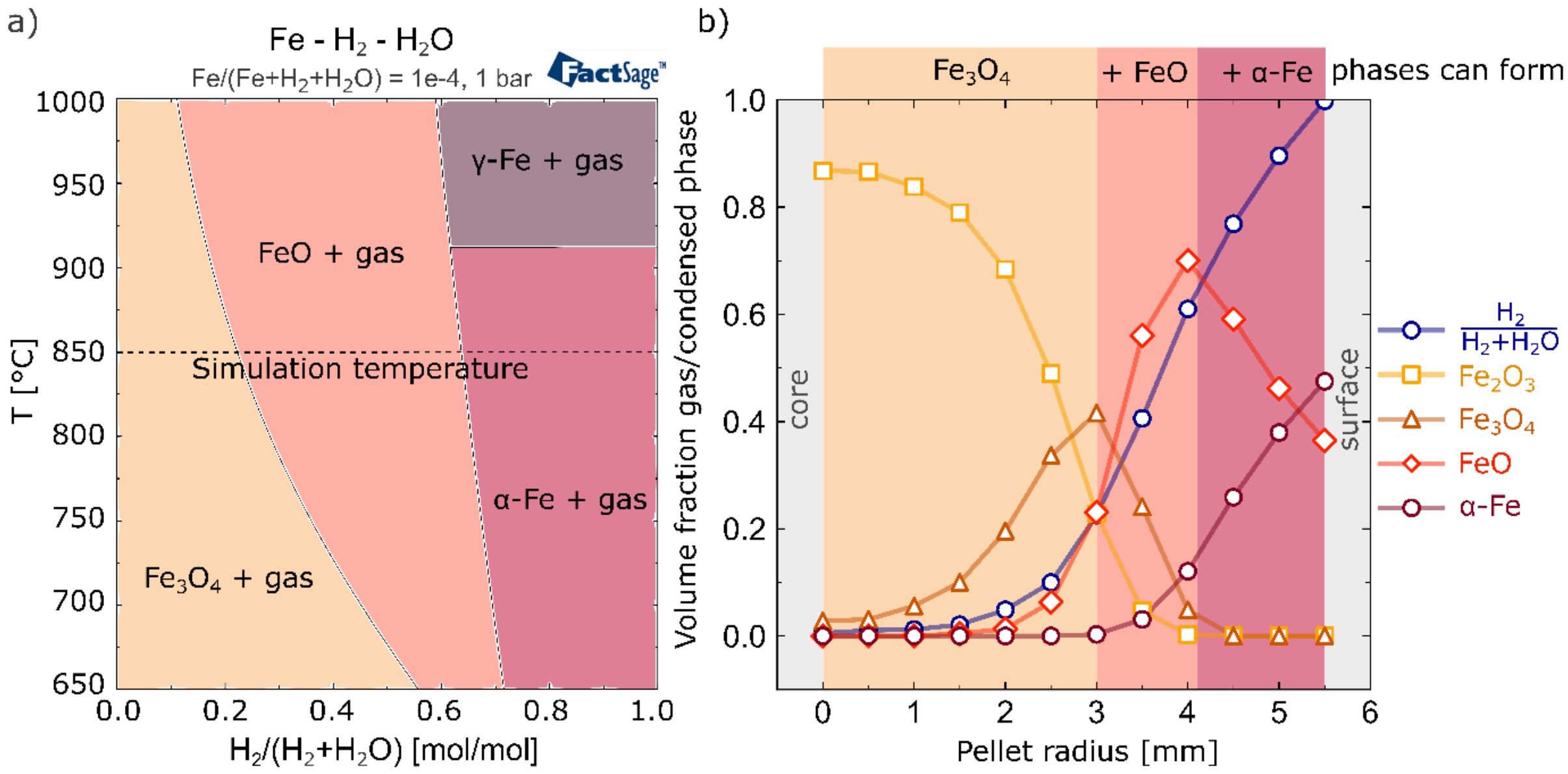

Figure 1: a) A Bauer-Glaessner type diagram of the *Fe-$H_2$-$H_2O$* system, where the different phase fields are colored to indicate where the different iron oxides and metallic iron are stable, calculated by *FactSage 8.3*. b) Predicted phase formation and $H_2$ concentration gradient based on a simulation employing the Porous Solid Model as reported by Fradet et al.[30] for $H_2$ reduction at 850°C of a hematite pellet, corresponding to 50% total conversion to *α-Fe*[30]. The simulation was conducted to match a conversion degree data set from Kazemi at al.[32] (Scenario 1). Based on the $H_2/(H_2+H_2O)$ ratio, the areas are colored in line with a) to indicate where the formation of $Fe_3O_4$, $FeO$, and metallic iron are thermodynamically possible.

In this study, we introduce a thermodynamic driving force $\Delta G$ in the source term, so that our model accounts for the local gas phase, using multi-phase, multi-component equilibria via the *CALPHAD* methodology (Gibbs free energy minimization). We derive a mathematical description for the reversible first-order rate of chemical reactions, where $\Delta G$ determines both, the extend and the direction of the reduction reaction. Besides ensuring thermodynamic consistency, this approach allows for reducing the number of fitting parameters: the reduction reactions are described independently of the reductant molecule, $CO$ or $H_2$, but are based only on the partial pressure of $O_2$ in the equilibrated gas phase. Additionally, homogeneous gas-gas reactions are not treated explicitly but gas species are assumed to be in equilibrium. Similarly to reduction, partitioning reactions are defined based on the activity of $C$ in the gas phase. Our model offers a framework for modeling microstructure evolution in pellet reduction processes. Unlike existing pellet reduction models, the model parameters are fitted not only using macroscopic reduction kinetics measured in a thermogravimetric analysis, but using reported radially resolved microstructure information of a pellet after interrupted reduction[33].

## 2. Methods

### 2.1. Chemical reactions in DRI and description of the thermodynamic driving force

Usually, the following reaction equations are used as basis to model the reduction of the different iron oxide phases in $H_2$- and $CO$-rich atmospheres[13,17,18,23,30,32,34–38]:

$$3\,Fe_2O_3 + H_2 \leftrightarrow 2\,Fe_3O_4 + H_2O \tag{1}$$

$$Fe_3O_4 + H_2 \leftrightarrow 3\,FeO + H_2O \tag{2}$$

$$FeO + H_2 \leftrightarrow Fe + H_2O \tag{3}$$

$$3\,Fe_2O_3 + CO \leftrightarrow 2\,Fe_3O_4 + CO_2 \tag{4}$$

$$Fe_3O_4 + CO \leftrightarrow 3\,FeO + CO_2 \tag{5}$$

$$FeO + CO \leftrightarrow Fe + CO_2. \tag{6}$$

For Equations (1)–(6), first-order reaction rate laws are applied, meaning that two fitting parameters, i.e., a pre-exponential factor and an activation energy, are introduced. Additionally, side reactions between the different gas species, e.g., methane reforming or the water-gas shift reaction, are sometimes ignored[18,30] and sometimes included[13,23]. Salucci et al. identified a significant scatter of the reported activation energies, e.g., for the magnetite to wüstite transition, the average activation energy based on 27 references is 66 kJ/mol with a standard deviation of 57.2 kJ/mol. It is concluded that "the current state of knowledge may not accurately capture the intricacies of the reduction process"[9]. Thus, it is reasonable to assume that the models published so far contain too many fitting parameters and consider too little experimental data to separate the effects of the different model parameters on sound thermodynamics grounds. Here, an approach is introduced to reduce the number of fitting parameters.

First, the six reaction equations implying independent behavior of $H_2$ and $CO$ as reductant are replaced by three reaction equations, focusing on the iron phase transformation. We assume that the rate-limiting step in reduction is the transport of oxygen from the oxide to the gas phase. This leads to:

$$3\,Fe_2O_3 \leftrightarrow 2\,Fe_3O_4 + [O]_{gas} \tag{7}$$

$$Fe_3O_4 \leftrightarrow 3\,FeO + [O]_{gas} \tag{8}$$

$$FeO \leftrightarrow Fe + [O]_{gas}. \tag{9}$$

Here, $[O]_{gas}$ refers to an oxygen atom in the gas phase, independent of the molecule in which the oxygen atom is bound, i.e., $H_2O$ or $CO_2$ in the current context. For each of these reactions demonstrated in Equations (7), (8), and (9) first-order reaction rate laws are applied with the modification that we multiply the first-order reaction rate with the driving force for the respective reduction reaction. This approach reduces the number of fitting parameters for the heterogeneous reactions by a factor of two. The mathematical approach is described in detail in Section 2.2.

Second, to calculate the driving force for the reduction, the partial pressure of oxygen is calculated using *ChemApp*[39,40] relying on thermodynamic data from *FactSage*[41,42] *8*.3 using the *GTOx* (2023) and *SGPS* (2023) databases for pure solid substances and the gas phase species, respectively, but other databases could have been used as well. An equilibrated gas phase is assumed which implies that homogeneous side-reactions in the gas phase are considered to be infinitely fast. The assumption of an equilibrated gas phase contrasts with the treatment by Takahashi et al.[34] which has also been adopted recently by Immonen and Powell for the water-gas shift reaction[23]. Their approach relies on four reaction rate constants for the water-gas shift reaction, since two different sets of forward and backward reaction rate constants are used depending on the degree of reduction, allegedly since metallic *Fe* and iron oxides act to a different extent as catalyst. Each of the reaction constants is described using a pre-exponential factor and an activation energy, leading to eight parameters fitted by Takahashi et al.[34] and supplemented by two more parameters describing the temperature dependence of the equilibrium constant. Takahashi et al.[34] reported experimentally measured gas compositions and temperatures in a lab-scale direct reduction furnace for a reduction experiment using a reductant gas mixture consisting of 71.5 vol.% $H_2$, 27.9 vol.% $CO$, and 0.6 vol.% $CH_4$[34]. Figure 2 compares the experimentally observed gas composition at different positions of the reactor with the gas composition predicted using equilibrium calculations in *FactSage*. In all areas that are not close to the gas inlet, the gas speciation is well captured using equilibrium calculations. Close to the gas inlet, where the concentration of $H_2$ and $CO$ is the highest, methane formation seems to be kinetically hindered. However, the water-gas shift reaction which is relevant for low amounts of $H_2$ and $CO$ can conveniently be modeled using an equilibrium assumption. Note that this approach already removes eight fit parameters introduced by Takahashi et al.[34] (and modified by Immonen and Powell[23]) to describe the water-gas shift reaction. Methane formation and reforming could be described as well by introducing virtual system components as constraints to the thermodynamic equilibrium[43], but this is not in the focus of the current work.

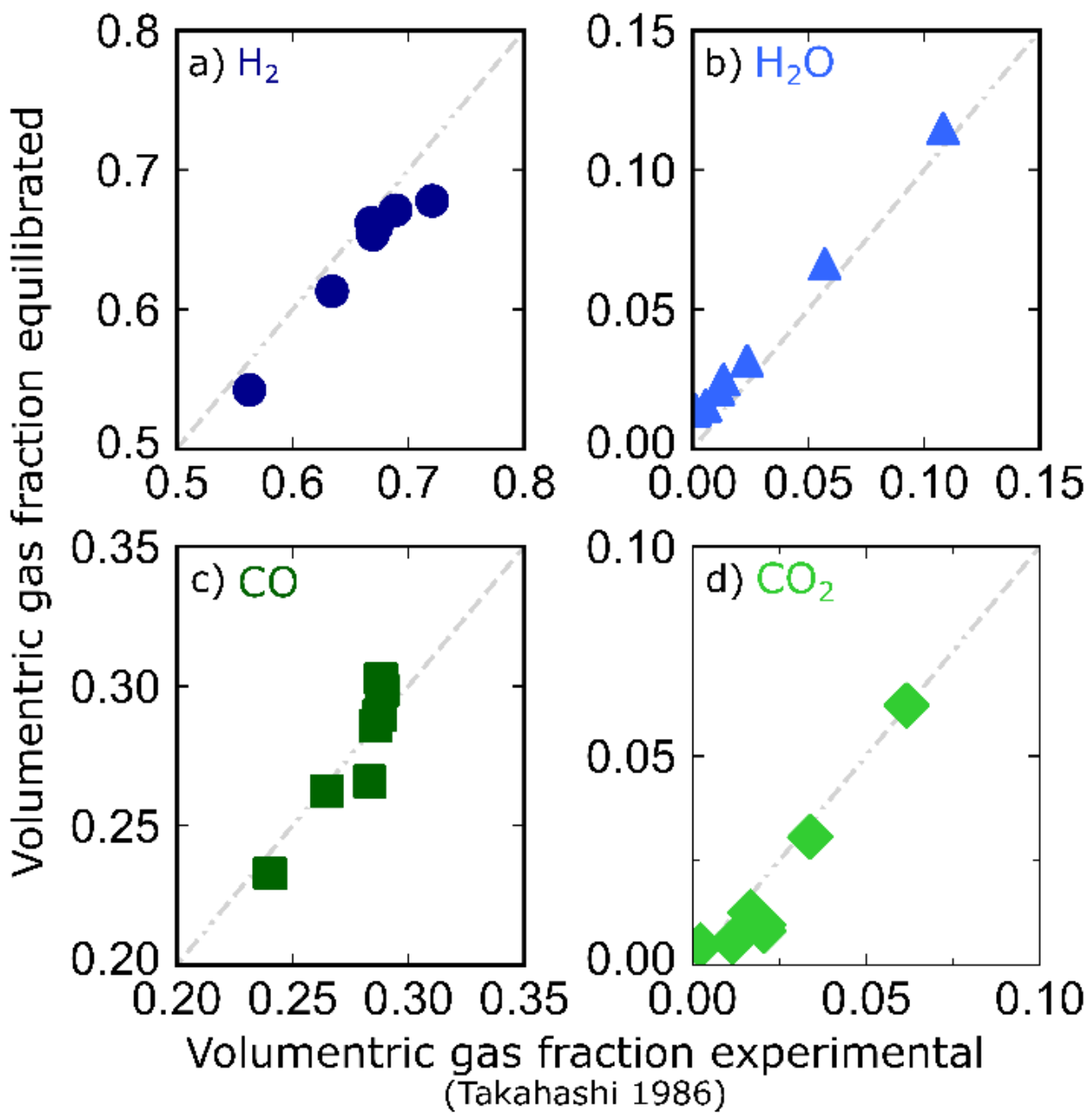

Figure 2: Volumetric gas fraction of a) $H_2$, b) $H_2O$, c) $CO$, and d) $CO_2$ experimentally measured at different locations in a lab-scale direct reduction furnace (x-axis) for a reduction experiment using a reductant gas mixture consisting of 71.5 vol.% $H_2$, 27.9 vol.% $CO$, and 0.6 vol.% $CH_4$ and compared to the respective calculated volumetric gas fractions in thermodynamic equilibrium of the gas mixture using *FactSage 8.3* (y-axis).

Additionally, two precipitation reactions are defined to consider the formation of cementite or graphite in Equation (10) and (11).

$$[C]_{gas} \leftrightarrow C \tag{10}$$

$$3\,Fe + [C]_{gas} \leftrightarrow Fe_3C \tag{11}$$

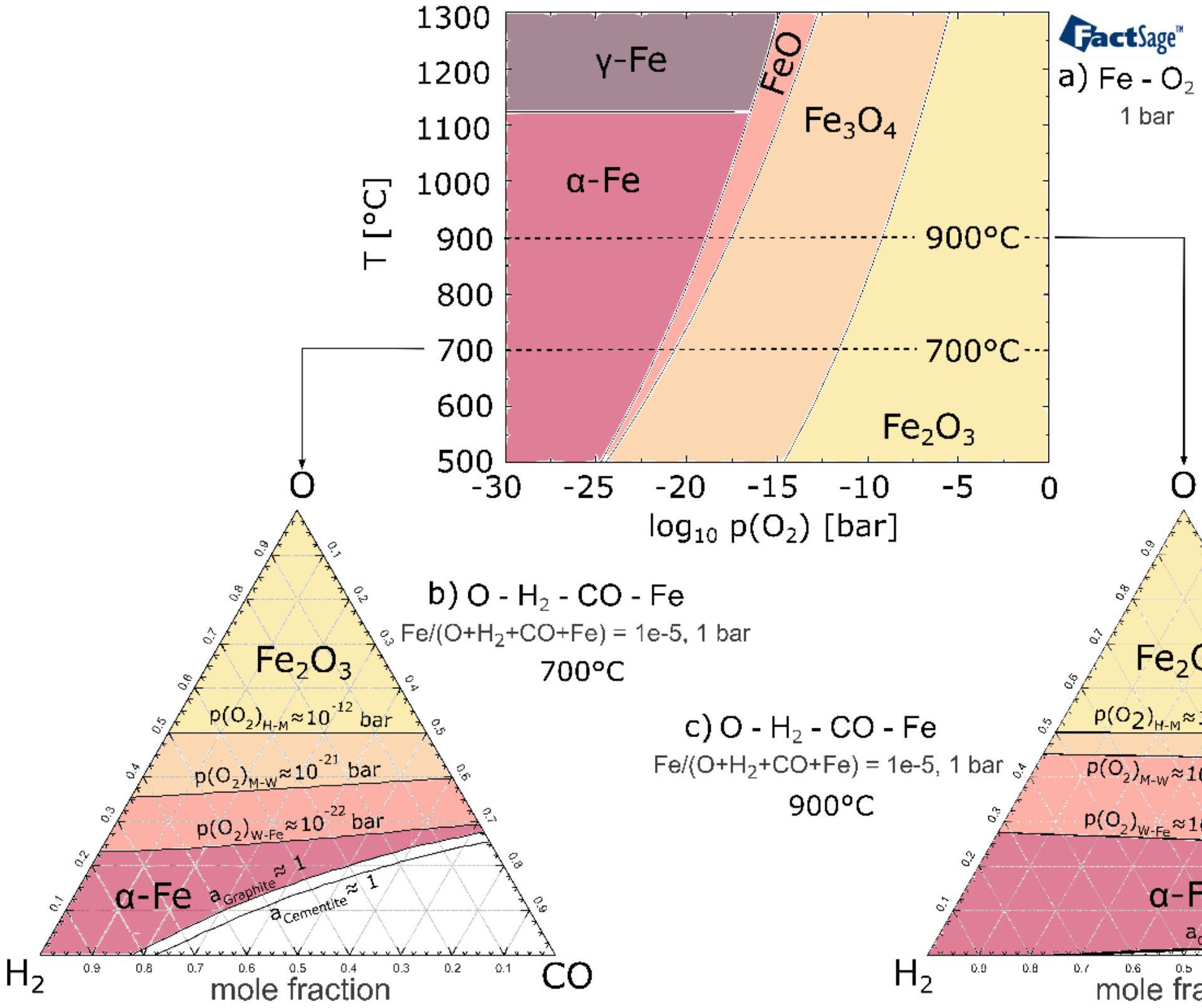

Figure 3: a) Fe-$O_2$ phase diagram at 1 bar for variable temperatures *T* and oxygen partial pressures *p($O_2$)* and pseudo-ternary phase diagrams of *O-$H_2$-CO-Fe* system at 1 bar with fixed amount of *Fe* at b) 700°C and c) 900°C. Thermodynamic phase stability regions are color-coded, with phase boundary *p($O_2$)* and activity *a* values indicated in b) and c). Phase diagrams are calculated by *FactSage 8.3*.

$[C]_{gas}$ here stands for a *C* atom that is present in the gas phase, no matter whether it is present in the form of *CO* or $CO_2$ molecules. Thus, all reactions considered are assumed to be reversible depending on the thermodynamic phase stabilities. At a given temperature and pressure, the thermodynamic stability regions of each solid phase are determined by the partial pressure of oxygen in the gas phase $p(O_2)_{gas}$, which is a function of the gas composition. This relationship is visualized in the *Fe-$O_2$* phase diagram shown in Figure 3a. The lines in the phase diagram mark the partial pressure necessary for a solid phase transformation to occur, denoted as $p(O_2)_{Phase\ 1-Phase\ 2,eq}$. From the ratio between the actual $p(O_2)_{gas}$ in the gas phase and $p(O_2)_{Phase\ 1-Phase\ 2,eq}$, the thermodynamic driving force $\Delta G$ for the solid phase transformation can be calculated using Equation (12).

$$\Delta G_{Phase1-Phase2} = R_{gas}\, T\, ln\left(\frac{p(O_2)_{Phase1-Phase2,eq}}{p(O_2)_{gas}}\right) \tag{12}$$

For partitioning reactions (see Equations (10) and (11)), the activity of *C* at equilibrium is used to calculate $\Delta G$, replacing the partial pressure of oxygen inside the logarithm term. The driving force for graphite precipitation has recently been demonstrated to be an excellent predictor for metal dusting attacks in *C-H-O-Ar*-containing atmospheres[44]. As the reduction process proceeds, reaction products such as $H_2O$ and $CO_2$ accumulate in the gas, altering the gas composition and thereby modifying $p(O_2)_{gas}$. These changes affect the stability regions of each solid phase, as shown in Figure 3b and c for 700 and 900°C, respectively, and consequently the reduction behavior of the iron oxide pellets.

### 2.2. Governing equations

In order to mathematically model the temporal evolution of gas species concentrations $C_i$ and solid-phase fractions $X_j$ anywhere inside the porous pellet, the mass conservation equation is simplified, neglecting convection but considering diffusion and a source term. In this approach, based on the work by Fradet et al.[45], the governing reactions for pellet reduction are formulated as Equation (13) for gas species and Equation (14) for solid phases.

$$\frac{\partial \varepsilon C_i}{\partial t} - \nabla\left(D_{eff,i} \nabla C\right) = S_i , \qquad i \in \{H_2, H_2O, CO, CO_2, CH_4\} \tag{13}$$

$$\frac{\partial (1-\varepsilon) X_j}{\partial t} = \frac{M_j}{\rho_j} S_j, \qquad j \in \{Fe_2O_3, Fe_3O_4, FeO, Fe, C, Fe_3C\} \tag{14}$$

In these equations, $\varepsilon, D_{eff,i}, M_j$, and $\rho_j$ represent porosity, the effective diffusion coefficient of gas species $i$, the molar mass, and the mass density of the solid phase $j$, respectively. Currently, all gas compounds $i$ consisting of *C*, *H*, or *O* are considered for the gas phase. The model is easily adaptable to include any other species as long as occurring gas-solid precipitation reactions are specified (e.g., Equation (10) and (11)). $S_i$ and $S_j$ refer to the source term for gas and solid species, respectively, and describe the net chemical consumption or production rate due to surface reactions.

Following the work of Fradet et al.[30], the partial differential equations in a spherical coordinate system with the radial coordinate $r$ are discretized in one dimension using *FDM*. By changing the mathematical formulation for the differential volume, it would also be possible to describe any other symmetry which might however require a higher dimensionality of the model, and thus more computational costs. The method of lines is applied, discretizing the spatial variables while keeping time continuous, transforming the system into a set of ordinary differential equations (*ODE*). In this work, the adaptive multistep *ODE* solver *LSODA* from the Python´s *SciPy* package[46] is used to solve the resulting system. By solving the governing equations for specific initial and boundary conditions, the reduction behavior of iron oxide pellets and its dependence on pellet properties and the reduction environment can be calculated. The interplay between the diffusion-driven transport and chemical reaction-driven source term provides insights into rate limiting steps and will be discussed in detail in the next sections.

#### *2.2.1. Diffusion term*

The diffusion term $\nabla\left(D_{eff,i} \nabla C\right)$ in the Equation (13) accounts for mass transport in the radial direction of the pellet, driven by concentration gradients, in accordance with Fick's second law of diffusion. Transport of multi-component gas mixtures in a porous medium differs from bulk diffusion[47] and can be described by the effective diffusivity term $D_{eff,i}$ for each gas species $i$, as defined in Equation (15).

$$D_{eff,i} = \varepsilon^{1.5} D_{int,i} \tag{15}$$

Here, $\varepsilon$ denotes the fraction of connected pores which commonly increases over the course of the reduction process and, hence, affects the effective diffusion of species. The exponent of 1.5 originates from the application of Bruggeman's model, which describes the theoretical relationship between tortuosity and porosity[48,49]. $D_{int}$ denotes the interparticle diffusion coefficient that accounts for the combined effects of the binary molecular diffusion and the Knudsen diffusivity. The latter particularly influences the diffusion behavior when the effective average pore diameter (*eAPD*) is small (i.e., in the range of 1 to 50 nm). In such cases, gas species frequently interact with the pore walls, influencing the overall diffusion behavior. Mathematical equations that describe $D_{int,i}$ are adapted from Metolina et al.[36] with slight modifications and are provided in detail in the Supporting Information (S1).

#### 2.2.2. Source term

We define a source term $S_{i/j}$ for each gas component *i* and each solid species *j* and it denotes the rate at which these species are consumed and produced due to chemical surface reactions within the pellet. Hence, each source term $S_{i/j}$ includes reaction terms $R_{phase\ 1-phase\ 2}$ for each reaction where *i* or *j* are consumed as reactants or formed as products. Table 1 provides the stochiometric relationships for the source term of each solid species at each spatial position *r*.

Table 1: Source term $S_j$ of solid species *j* at each spatial position *r* with specified numerical description of source terms of each reaction where species *j* is either consumed or produced in a certain molar amount. *[C]* refers to carbon species in the gas phase, whereas *C* refers to carbon in the solid state.

| Source term of the condensed phase [mol s$^{-1}$ m$^{-3}$] | Numerical description |
|---|---|
| $S^r_{hematite}$ | $-\,3\,R^r_{H-M}$ |
| $S^r_{magnetite}$ | $2\,R^r_{H-M} - R^r_{M-W}$ |
| $S^r_{wüstite}$ | $3\,R^r_{M-W} - R^r_{W-Fe}$ |
| $S^r_{iron}$ | $R^r_{W-Fe} - 3\,R^r_{Fe-Fe3C}$ |
| $S^r_{graphite}$ | $R^r_{[C]-C}$ |
| $S^r_{cementite}$ | $R^r_{Fe-Fe3C}$ |

The evolution of gas species is tracked by the *H*, *O*, and *C* components, as shown in Table 2, because the gas phase is always equilibrated. Since there is no reaction defined that transforms hydrogen to solid state, the corresponding source term $S^r_{[H]}$ is zero for all *r*. This approach reduces the number of dependent variables to be iterated over by the solver from five to three for a *H-O-C* system. The source term for each gas species ($H_2, H_2O, CO, CO_2, CH_4$) is then derived from the source terms of the individual gas components *H-O-C*, considering the thermochemical equilibria.

Table 2: Source terms of gas species components *H, O,* and *C* at each spatial position *r* with specified numerical description of source terms of each reaction where the components are either consumed or produced in a certain molar amount. *[C]* refers to carbon species in the gas phase, whereas *C* refers to carbon in the solid state.

| Source term of the gas components [mol s$^{-1}$ m$^{-3}$] | Numerical description |
|---|---|
| $S^r_{[H]}$ | 0 |
| $S^r_{[O]}$ | $R^r_{H-M} + R^r_{M-W} + R^r_{W-Fe}$ |
| $S^r_{[C]}$ | $-\,R^r_{[C]-C} - R^r_{Fe-Fe3C}$ |

The reaction term $R_{phase\ 1-phase\ 2}$ is then defined as:

$$R_{phase\ 1-phase\ 2} = k_{phase\ 1-phase\ 2} * C_{i,react} * X_{j,react} * \Delta G_{phase\ 1-phase\ 2} \tag{16}$$

Equation (16) is the main mathematical description to define each chemical reaction, where $C_{i,react}$ and $X_{j,react}$ denote the concentration of reacting gas species and the mole fraction of the reacting solid species $j$ ( $phase\ 1$), respectively. $C_{i,react}$ for gas mixtures refers to the sum of reacting gas, i.e., $C_{H_2} + C_{CO}$ when reducing, and $C_{H_2O} + C_{CO_2}$when oxidizing. The multiplication of $C_{i,react}$ and $X_{j,react}$ describes the probability of reduction of solid species $j$. $k_{phase\ 1-phase\ 2}$ denotes the kinetic rate constant of the chemical reaction and is included to capture kinetic limitations of the reaction. Finally, $\Delta G_{phase\ 1-phase\ 2}$ is included to dynamically evaluate the thermodynamic feasibility of each reaction step, ensuring that the model is informed by the equilibrium state and driving forces under the specified conditions. The extent and direction of the reactions are governed by $\Delta G$.

### 2.3. Input parameters for pellet reduction model

Based on the governing equations, the temporal evolution of species at discrete locations in the porous pellet can be evaluated during reduction and can be compared to experimental results. In this work, experimental data, mainly experimental pellet conversion degrees, from three different scenarios is evaluated, each defined by a distinct set of input parameters as shown in Table 3.

Table 3: Input parameters for the pellet reduction model for each simulation discussed in Section 3.

| Parameters | Scenario 1 | Scenario 2 | Scenario 3 |
|---|---|---|---|
| Temperature [°C] | 850°C | 850°C | 700°C |
| Pressure [Pa] | $10^5$ | $10^5$ | $10^5$ |
| Pellet diameter [mm] | 11 | 11 | 11 |
| Initial porosity [%] | 26 | 27 | 24 |
| Initial phases [vol.%] | 96 $Fe_2O_3$, 4 gangue | 96 $Fe_2O_3$, 4 gangue | 96 $Fe_2O_3$, 4 gangue |
| Boundary gas species [vol.%] | 100 $H_2$ | 60 $H_2$ – 40 CO | 100 $H_2$ |
| Reference | [32] | [50] | [33] |

In addition to these input parameters, several material-specific properties, such as molar masses, densities, and diffusion volumes, were included in the simulations. The complete list of input parameters is provided in the Supporting Information (S2). The *eAPD* and the kinetic parameters $k_{H\text{-}M}$, $k_{M\text{-}W}$, and $k_{W\text{-}Fe}$ for pure $H_2$ reduction as well as two additional parameters for the *C*-precipitation reactions in case of mixed gas atmosphere were fitted for each scenario individually to match the experimental results. Optimal fits to convergence degree data were determined based on the root-mean-square metric. As described in Section 3.3, we additionally suggest that radially resolved phase evolution should be considered when determining kinetic constants.

### 2.4. Calculation of conversion degree from hematite to α-iron

The experimental conversion degree is a quantity derived from the mass loss of iron oxide pellets during isothermal reduction, calculated by the standard gravimetric method[50,51]. Here it is referred to as the global conversion degree which assumes a holistic consideration of the total pellet mass. A conversion degree of 0 corresponds to a pellet under initial conditions whereas a conversion degree of 1 indicates approximately 100% metallic *α-Fe*, omitting the mass of unreacted gangue. To simulate the mass loss in the model, the local phase evolutions within the pellet are used like in the work by Fradet et al.[30]. Consequently, the local conversion $\Phi_r(t)$ at any radial point *z* and time *t* is calculated by:

$$\Phi_r(t) = \frac{w_0 - w_r(t)}{w_0 - w_\infty} \tag{17}$$

Here, $w_0$, $w_r(t)$, and $w_\infty$ are the initial solid mass (per unit volume of pellet) at t = 0, the local mass (per unit volume of pellet) at radial point *r*, and the theoretically reduced mass after complete reduction to metallic *α-Fe*, respectively. A global conversion degree $F(t)$ is then obtained by integrating $\Phi(r,t)$ over the entire volume of the pellet $V_{pellet}$ with radius $r_{pellet}$.

$$F(t) = \frac{1}{V_{pellet}} \int_0^{r_{pellet}} \Phi(r,t)[4\,\pi\,r^2]dr. \tag{18}$$

The global conversion degree for each simulation is calculated using Equation (18) and subsequently fitted to the experimentally obtained conversion degree.

## 3. Results and discussion

### 3.1. Demonstration of thermodynamic consistency for $H_2$ as reducing gas

In Figure 4a, a simulated conversion degree curve, based on input parameters fitted to match the experimental conversion degree data reported by Kazemi et al.[32] (Scenario 1), is displayed. The simulation corresponding to the data presented in Figure 1b is based on the same Scenario 1 data set which also reported excellent agreement with the experimentally measured conversion degree curve[30,32], however leading to thermodynamically inconsistent results as discussed in the introduction.

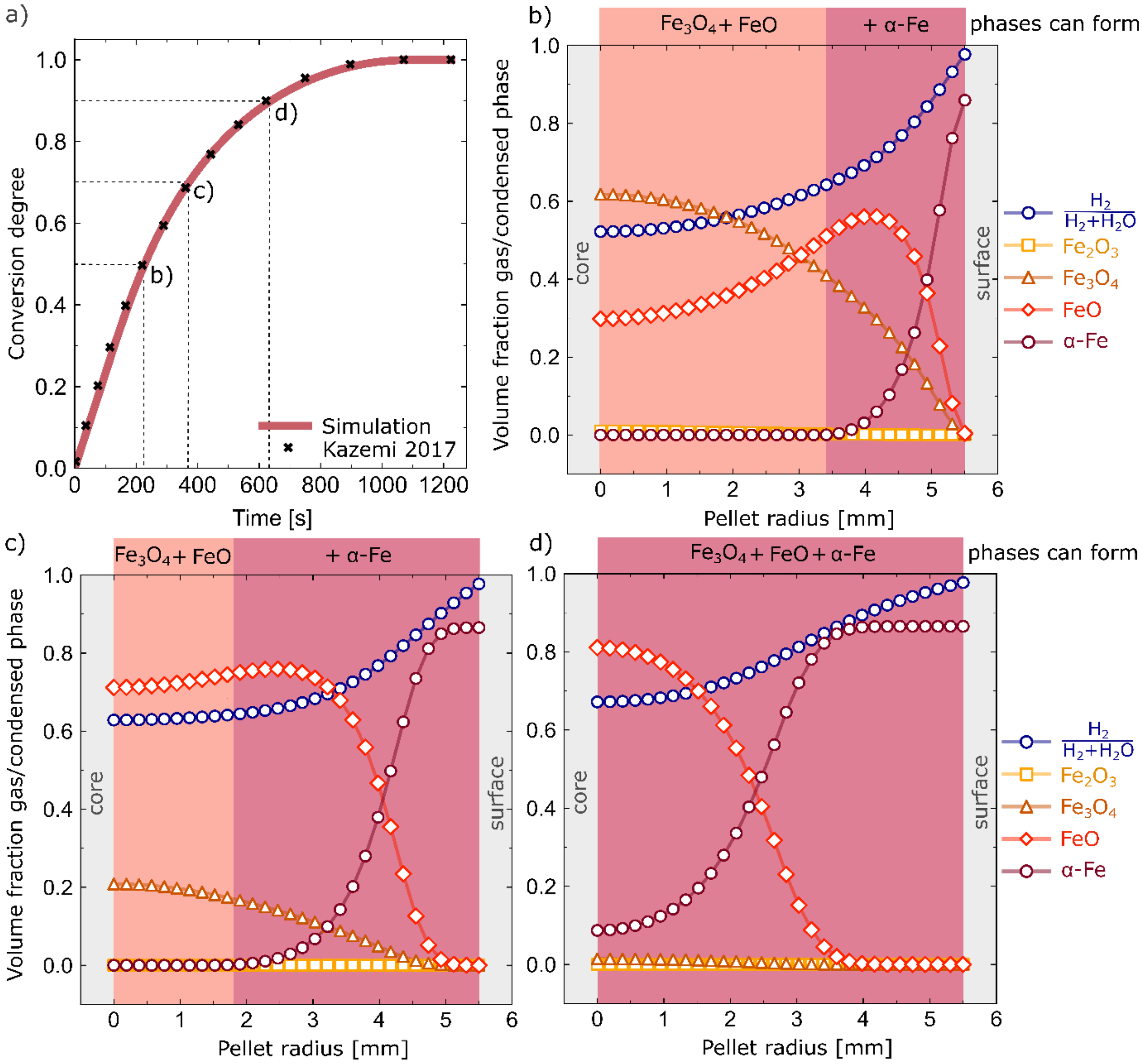

Figure 4: a) Experimental pellet reduction data (black x) from Kazemi at al.[52] (Scenario 1) and simulated conversion degree curve (red solid line) employing the modified porous solid model with input parameters $k_{H\text{-}M}$ = 1.29E-04 mol J$^{-1}$ s$^{-1}$, $k_{M\text{-}W}$ = 3.53E-04 mol J$^{-1}$ s$^{-1}$, $k_{W\text{-}Fe}$ = 2.68E-03 mol J$^{-1}$ s$^{-1}$, and effective average pore diameter of 100 nm, fitted to match the experimental data. b)-d) Based on the simulation, predicted mole fraction of gas/condensed phases over pellet radius for a conversion of 50 %, 75 %, and 95 %, respectively. Regions of thermodynamically stable phase formation based on the phase diagram in Figure 1a are marked in the background.

After approximately 225 s, 50 % conversion of the pellet is reached. The simulated radial distribution of gas species and condensed phases at this state are displayed in Figure 4b. In Figure 4c and d the phase evolution for 75 % and 95 % global conversion are shown, respectively. A small amount of $O_2$ is produced but is not shown due to its relatively small molar fractions (<$10^{-10}$ mol%) within the gas phase. Similar to the simulation in Figure 1b, the model predicts the outer region of the pellet to be exposed to a higher $H_2/(H_2 + H_2O)$ ratio than the pellet core. The $H_2/(H_2 + H_2O)$ ratio is strongly related to local $p(O_2)_{gas}$, which, as displayed in Figure 1a, determines the thermodynamic stability regimes of *α-Fe*, *FeO*, and $Fe_3O_4$ (shown as background colors) as $p(O_2)_{H-M,eq}$, $p(O_2)_{M-W,eq}$, and $p(O_2)_{W-Fe,eq}$ are constant under isothermal conditions. Based on the simulation with our model, *α-Fe* formation is observed only within the respective thermodynamic stability region of *α-Fe* where $p(O_2)_{gas}$ is low enough to stabilize metallic iron, i.e., for a 50 % conversion degree it is observed at distances of more than 3.42 mm from the core. For this global conversion degree, magnetite dominates in core regions where the reducing potential of the gas is weaker, while wüstite shows the highest molar fraction near mid-radius, transitioning to metallic iron at the surface. Consequently, a smooth and continuous phase evolution across the pellet radius is observed. This evolution is influenced by the local chemical potentials set by the components in the gas, where phases can only occur at conditions where the thermodynamic stability criterion is met.

### 3.2. Implications for a $H_2$-CO reducing gas mixture

Homogeneous gas-gas reactions, e.g., the water-gas shift reaction, are accounted for by the assumption of an equilibrated gas-phase rather than by introducing additional fitting parameters for each reaction. Consequently, the model remains adaptable to consider diverse types of gas mixtures encountered in direct reduction operations. When *C*-bearing gases are involved, carbides or graphite may form and partition among the various phases, potentially rendering some of the *C* into a solid state which is an important and beneficial effect for future sustainability considerations when using mixed reductants. A parametrization study to assess the sensitivity and extrapolation behavior of our model under industrially relevant conditions, starting from a baseline corresponding to Scenario 2 (750°C ,$10^5$ Pa, and 60 $H_2$-40 CO), can be found in the Supporting Information (S3), including conversion degree curves in Figure S1. A simulation of the conversion degree (Figure 5a) and temporal porosity and phase evolution during the reduction (Figure 5b) are displayed based on a fit to experimental conversion degree data of pellets during a reduction in a 60/40 vol.% $H_2/CO$ mixed gas atmosphere, published by Kazemi et al.[50] (Scenario 2). The local evolution of the condensed phases (Figure 5c and d), gas phase species (Figure 5e), and the porosity (Figure 5f) over pellet radius for a pre-reduced pellet (after 1400 s) are further illustrated.

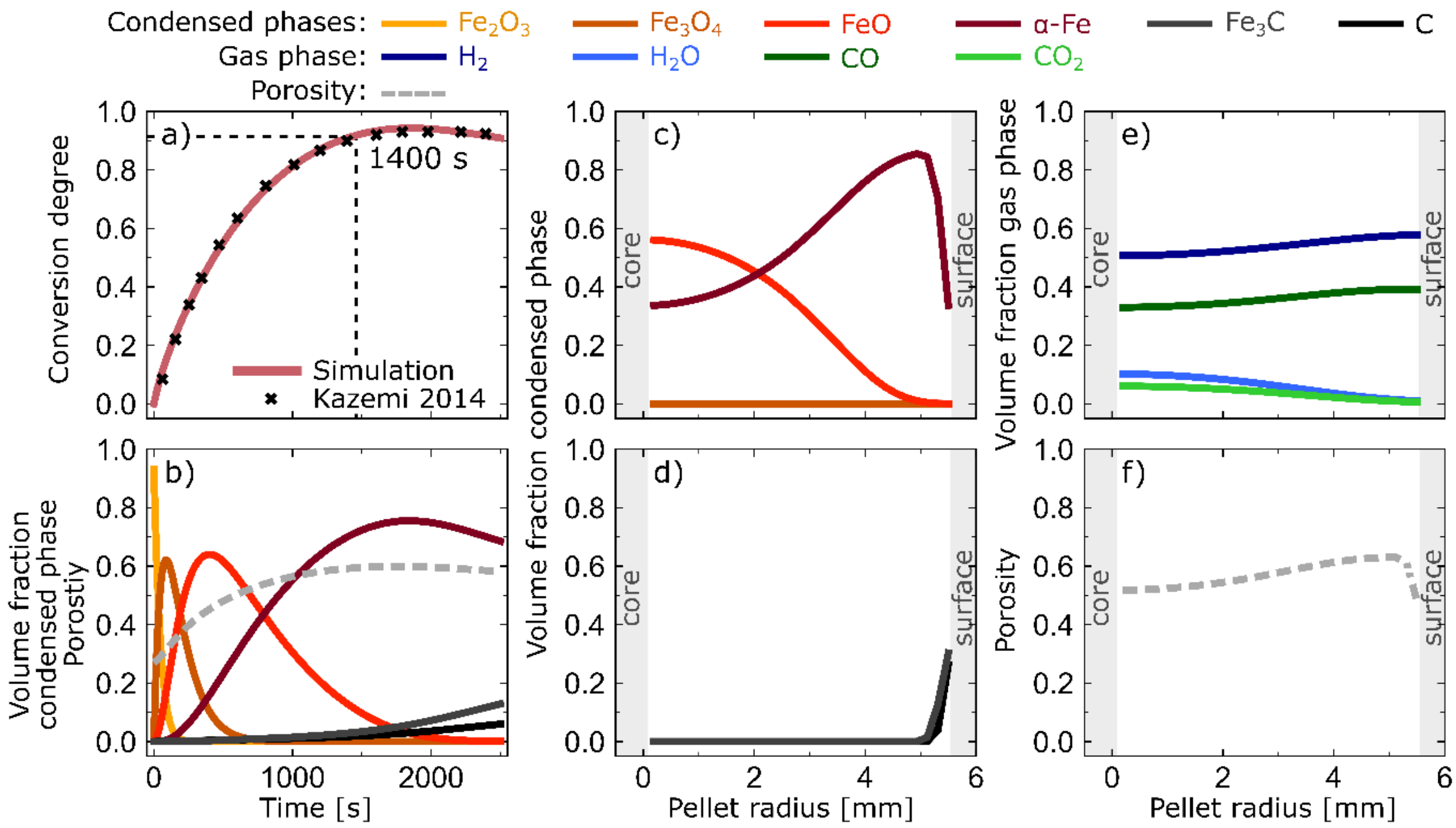

Figure 5: a) Experimental pellet reduction data (black x) from Kazemi at al.[50] (Scenario 2) and simulated conversion degree curve (red solid line) employing the modified porous solid model with input parameters $k_{H\text{-}M}$ = 1.23E-04 mol $J^{-1}$ $s^{-1}$, $k_{M\text{-}W}$ = 2.57E-04 mol $J^{-1}$ $s^{-1}$, $k_{W\text{-}Fe}$ = 3.11E-04 mol $J^{-1}$ $s^{-1}$, $k_{[C]\text{-}C}$ = 1.02E-03 mol $J^{-1}$ $s^{-1}$, $k_{Fe\text{-}Fe3C}$ = 5.57E-04 mol $J^{-1}$ $s^{-1}$, and effective average pore diameter of 100 nm, fitted to match the experimental data. Based on the simulation, b) the evolution of porosity (grey dashed line) and condensed phase fractions over time where porosity represents the free volume fraction for gas transport, while condensed phases occupy the remaining space, ensuring their sum is always 1 at each timestep, c) simulated volume fraction of *α-Fe* and *Fe*-oxides, d) simulated volume fraction of graphite (*C*) and cementite ($Fe_3C$), e) simulated volume fraction of gas species, f) simulated porosity. Results c)–f) are shown at 1400 s of reduction across the radius of the pellet.

The temporal evolution of porosity (Figure 5b) can be related to the phase evolution and is governed by mass loss, which occurs due to oxygen removal as $Fe_2O_3$ reduces to *Fe*, adding to the inherited porosity of the pellet at the initial stage. Pellet shrinkage as well as structural changes, like local cracks and delamination observed in *FeO*-to-*Fe* transformations[25], are not accounted for in the current model. These factors may significantly influence porosity and gas transport behavior in practical applications and will be implemented at a later point in time. Furthermore, precipitates like *C* or $Fe_3C$ are known to reduce porosity by occupying and obstructing the pore networks, especially in areas where these phases accumulate within the pellet[37,53]. The formation of *C* and $Fe_3C$ is observed at later stages of reduction (Figure 5b), i.e., from 1000 s, and at the surface of the pellet (Figure 5d), and corroborates well with the microstructural observations by Kazemi et al.[37]. It is evident that the *C* and $Fe_3C$ formation in the near surface region locally decreases porosity from 0.62 at 5 mm to 0.48 at 5.5 mm from the core (Figure 5f). As the reaction advances to 1800 s, *α-Fe* becomes prominent, and ultimately transforms into $Fe_3C$ towards the surface of the pellet. Carbon deposition is promoted by the gas composition with high carbon activity in the gas phase (Figure 5e). The reducing potential of the gas, i.e., $H_2 + CO/(H_2 + CO + H_2O + CO_2)$ varies over time and over the pellet radius, due to the dynamic nature of gas transport and chemical consumption that influences the local variations of phase transformations, and thus the overall reaction kinetics.

It should be noted that the consistency of the model behavior and the limited, and still rather qualitative microstructural data are not yet seen as a fully sufficient corroboration that the physical mechanisms are described appropriately with the current model. A strong agreement between the model fit and experimental data, as observed in Figure 4a and Figure 5a, is achieved by fine-tuning the unknown fitting parameters ($k_{H-M}$, $k_{M-W}$, $k_{W-Fe}$, $k_{[C]-C}$, $k_{Fe-Fe3C}$, *eAPD*). However, multiple sets of these six parameters can yield similar fits to the conversion curve. While this is usually not addressed in publications on pellet reduction models, it is reasonable to assume that the other pellet reduction models suffer from the same weakness, given the fact that they contain an even larger set of fitting parameters.

To overcome the problem of fitting kinetic parameters in underdetermined systems, it is suggested here to use microstructure information of interrupted reduction experiments. This is explored in the following section.

### 3.3. Microstructurally informed fitting of kinetic parameters

The challenge of fitting kinetic parameters in pellet reduction models underscores the need for additional experimental data, such as spatially resolved measurements at pre-reduced states. As an example, six similar fits employing our model of the experimentally determined conversion degree curves published by Ma et al.[33] using very different parameter sets are shown in Figure 6a. Note that these six curves all overlap (i.e., the simulated data are in good correlation with the experimental mass loss presented by the conversion degree). The corresponding fitted parameter sets are presented in Table 4, indicating that the conversion degree curve can be fitted while assuming *eAPDs* ranging two orders of magnitude between 30 and 500 nm. It can be seen that an increase in *eAPD* can be compensated for by a decrease in $k_{W-Fe}$ (i.e., the reaction rate of the transformation of wüstite to metallic iron). Figure 6b shows the corresponding simulated $H_2/(H_2+H_2O)$ ratios along the radius after 1800 s of reduction. It is noticeable that for *eAPDs* below 100 nm a distinct $H_2/(H_2+H_2O)$ gradient exists, indicating that the reaction progress is diffusion limited. For *eAPD* larger than 100 nm, the concentration gradient in the gas phase is largely absent. All parameter sets yield distinctly different spatial evolutions of the condensed phases, as shown in Figure 6c–e, where the simulated line plots are compared with data from experimentally observed local phase analysis after 1800 s of reduction.

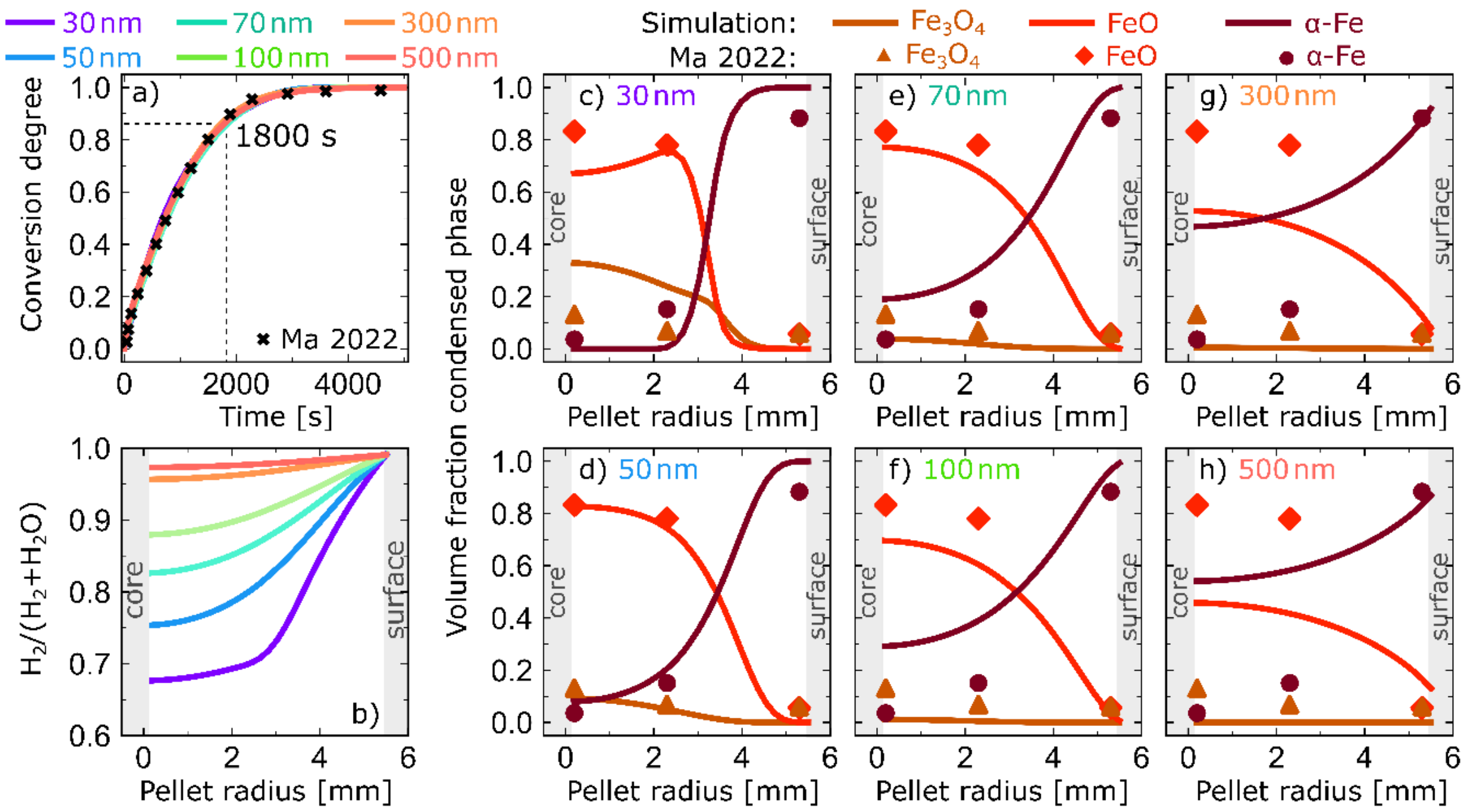

Figure 6: a) Experimental pellet reduction data (black x) from Ma at al.[52] (Scenario 3) and simulated conversion degree curves (solid lines) employing the modified porous solid model for input parameter sets (p1–p6) with varying effective average pore diameters displayed in Table 4, and fitted to match the experimental data. b) The simulated $H_2/(H_2+H_2O)$ ratios at 1800 s of reduction plotted over the pellet radius. c)–h) Based on the simulation, condensed phase evolution over radius (lines) plotted together with experimental x-ray diffraction data, reported by Ma at al.[33], on $Fe_3O_4$ (triangular shape), *FeO* (diamond shape), and *α-Fe* (circle) content at different locations inside the pellet after 1800 s of reduction for simulations with varying effective average pore diameter.

For *eAPDs* below 30 nm, low-residual fits to the experimental conversion degree of the Scenario 3 are impossible with adjusting the kinetic constants only because the system becomes excessively diffusion controlled. For small *eAPDs* of 30 and 50 nm, the limit for hydrogen diffusion to the pellet core causes a pronounced gradient in formed phases from the core to the surface. For large *eAPD*s such as 500 nm, the gradient in spatially resolved phase formation almost disappears, the reduction kinetics of the pellet depend mainly on the reaction rates of the three reduction reactions (Equation (7), (8), and (9)) rather than on diffusion. The best agreement between simulated and experimentally observed condensed phase fraction was achieved for an *eAPD* of 50 nm in parameter set p2. These results demonstrate that spatially resolved data can and should be used to avoid overfitting of parameters in direct reduction models, since conversion data alone is insufficient for validating pellet reduction models.

Table 4: Parameter sets for the four fitting parameters: effective average pore diameter (*eAPD*), kinetic constants for hematite-magnetite ($k_{H-M}$), magnetite-wüstite ($k_{M-W}$), and wüstite-iron ($k_{W-Fe}$) which yield the simulations displayed in Figure 6.

| Parameter set | eAPD [nm] | $k_{H-M}$ [mol J$^{-1}$ s$^{-1}$] | $k_{M-W}$ [mol J$^{-1}$ s$^{-1}$] | $k_{W-Fe}$ [mol J$^{-1}$ s$^{-1}$] |
|---|---|---|---|---|
| p1 | 30 | 7.42E-05 | 1.17E-04 | 3.80E-03 |
| p2 | 50 | 7.42E-05 | 1.42E-04 | 3.65E-04 |
| p3 | 70 | 7.42E-05 | 1.30E-04 | 2.04E-04 |
| p4 | 100 | 7.42E-05 | 1.30E-04 | 1.54E-04 |
| p5 | 300 | 7.04E-05 | 8.90E-05 | 9.89E-05 |
| p6 | 500 | 6.80E-05 | 8.65E-05 | 8.03E-05 |

While the model presented here is able to not only reproduce conversion degree curves but also spatially resolved phase formation, a number of extensions are still required. These include coverage of gangue components, inclusion of nucleation and growth as well as solid state diffusion, treatment of variations in pellet size and shape, uncertainties in the state of porosity, and delamination or cracking during experiments. In this study, we assume all pores to remain fully connected and use a single effective average pore diameter, although earlier research[54–57] indicates that pores can be isolated, and pore dimensions evolve both spatially and temporally. Further investigations into percolation processes, from both modeling and experimental perspectives, are needed to align model predictions with observed pellet behavior. It is hoped that the progress in pellet reduction models reported here inspires extensive investigations of the spatially resolved microstructures of pellets (e.g., using x-ray diffraction or scanning electron microscopy and energy dispersive x-ray spectroscopy or electron backscatter diffraction) in intermediate states of reduction.

## 4. Conclusions

In this study, we presented a numerical model for simulating the direct reduction of iron oxide pellets with mixed gas feedstock, an extension to the porous solid model. We propose the introduction of a driving force term to reaction equations to ensure thermodynamic consistency which is shown to be one critical weakness of existing models. Within our approach, only few kinetic constants and potentially the effective average pore diameter of the pellet need to be fitted. Fitting solely to the experimentally determined pellet conversion degree, as is practice in state-of-the-art pellet reduction models, is shown to be an underdetermined approach. By specifying general reduction equations for each iron oxide reactant, but not for each reducing agent separately, we reduce the number of kinetic fitting parameters from six to three for $H_2/CO$ gas mixtures. Furthermore, by coupling the chemical reactions to a thermochemical solver that determines the overall equilibrium composition of the gas phase, the model does not incorporate predefined equilibrium constants for gas-gas reactions; instead, it relies on the real-time (position and time dependent) thermodynamic state of the gas phase via the local $O_2$ partial pressure, to dictate how the iron oxides reduce. This approach eliminates the need for fitting kinetic parameters for gas-gas reactions.

The use of this pellet-scale model is demonstrated for reduction scenarios in $H_2$ and mixed $H_2$-$CO$ reducing gas atmospheres. The findings highlight that incorporating thermodynamics into pellet-reduction models is critical for determining the extent and reversibility of the chemical reactions. Gas diffusion is found to play an important role in reaction kinetics, emphasizing the need for further in-operando or interrupted characterization of pellet properties, especially porosity and pore size. These structural properties are essential for defining realistic boundary conditions, which remain necessary even in this extended porous solid model with fewer fitting parameters. While the strength of the proposed model lies in its thermochemical foundation, it does not currently account for nucleation and growth mechanisms, assumes an isothermal process and that all porosity is connected and neglects certain structural defects that develop during the reduction process. Further improvements can be achieved by integrating a heat balance as well as experimental data on porosity, pellet size, and microstructural changes at various stages of reduction and at different temperatures, thereby refining and validating the model's parameterization.

The reported modeling framework provides a mean-value approximation approach to predict the reduction kinetics and microstructural evolution of iron oxide pellets over pellet radius. This work lays the foundation for physically sound furnace-scale simulations that accurately capture the complexities of iron ore reduction processes and enable to identify pathways toward the decarbonization of the steel metallurgy.

**Acknowledgements**

The authors acknowledge funding through the EFRE-JTF ProDekan project (EFRE-20400099) by the state of North Rhine-Westphalia and co-financed by the European Union. This work was carried out as part of the International Max Planck Research School for Sustainable Metallurgy (IMPRS SusMet).

**Conflict of Interest Statement**

The authors disclose that ÖKB, FY, ALR, and MtB work for GTT-Technologies, while MtB is minority shareholder of GTT-Technologies. GTT-Technologies is developing and marketing the commercial API ChemApp and is co-developing and marketing the integrated thermodynamic software package FactSage which have been used here. GTT-Technologies also aims to commercialize the software presented here.

**Data Availability Statement**

The data that support the findings of this study are available from the corresponding author upon reasonable request.